\documentclass[a4paper,11pt]{article}
\usepackage{pos}
\usepackage{xspace}

\title{
  Impact of SeaQuest data on sea-quark PDFs at large $x$
}
\ShortTitle{
  Impact of SeaQuest data on sea-quark PDFs at large $x$
}

\author[a]{Sergey Alekhin}
\author*[a]{Maria Vittoria Garzelli}
\author[b]{Sergey Kulagin}
\author[a]{Sven-Olaf Moch}

\affiliation[a]{Universit\"at Hamburg, {II}. Institute for Theoretical Physics,\\
Luruper Chaussee 149, 22761 Hamburg, Germany}
\affiliation[b]{Institute for Nuclear Research of the Russian Academy of Sciences \\ 117312 Moscow, Russia}

\emailAdd{sergey.alekhin@desy.de}
\emailAdd{maria.vittoria.garzelli@desy.de}
\emailAdd{sergey.kulagin@gmail.com}
\emailAdd{sven-olaf.moch@desy.de}

\abstract{ 
  We investigate the impact of the recently released FNAL-E906 (SeaQuest) data on the ratio of proton-deuteron and proton-proton Drell-Yan production cross-sections on the sea quark PDFs. We find that they have constraining power on the light-quark sea isospin asymmetry $(\bar{d}-\bar{u})(x)$ and on the $(\bar{d}/\bar{u})(x)$ ratio at large longitudinal momentum fraction $x$ values,
  and that their constraints turn out to be compatible with those from Drell-Yan data in collider experiments (Tevatron and Large Hadron Collider) and in the old fixed-target experiment by the
  FNAL-E866 collaboration. We study the impact of nuclear corrections due to the deuteron target, finding them within 1\% in most of the kinematic region covered by SeaQuest. We perform a new proton PDF fit, including SeaQuest data, using the ABMP16 methodology and we compare it to the ABMP16 baseline.
}

\FullConference{The European Physical Society Conference on High Energy Physics (EPS-HEP2023)\\
 21-25 August 2023\\
Hamburg, Germany\\}

\begin{document}
\maketitle

\section{Introduction}
\label{intro}

\begin{figure}[h]
\centerline{\includegraphics[width=0.62\textwidth]{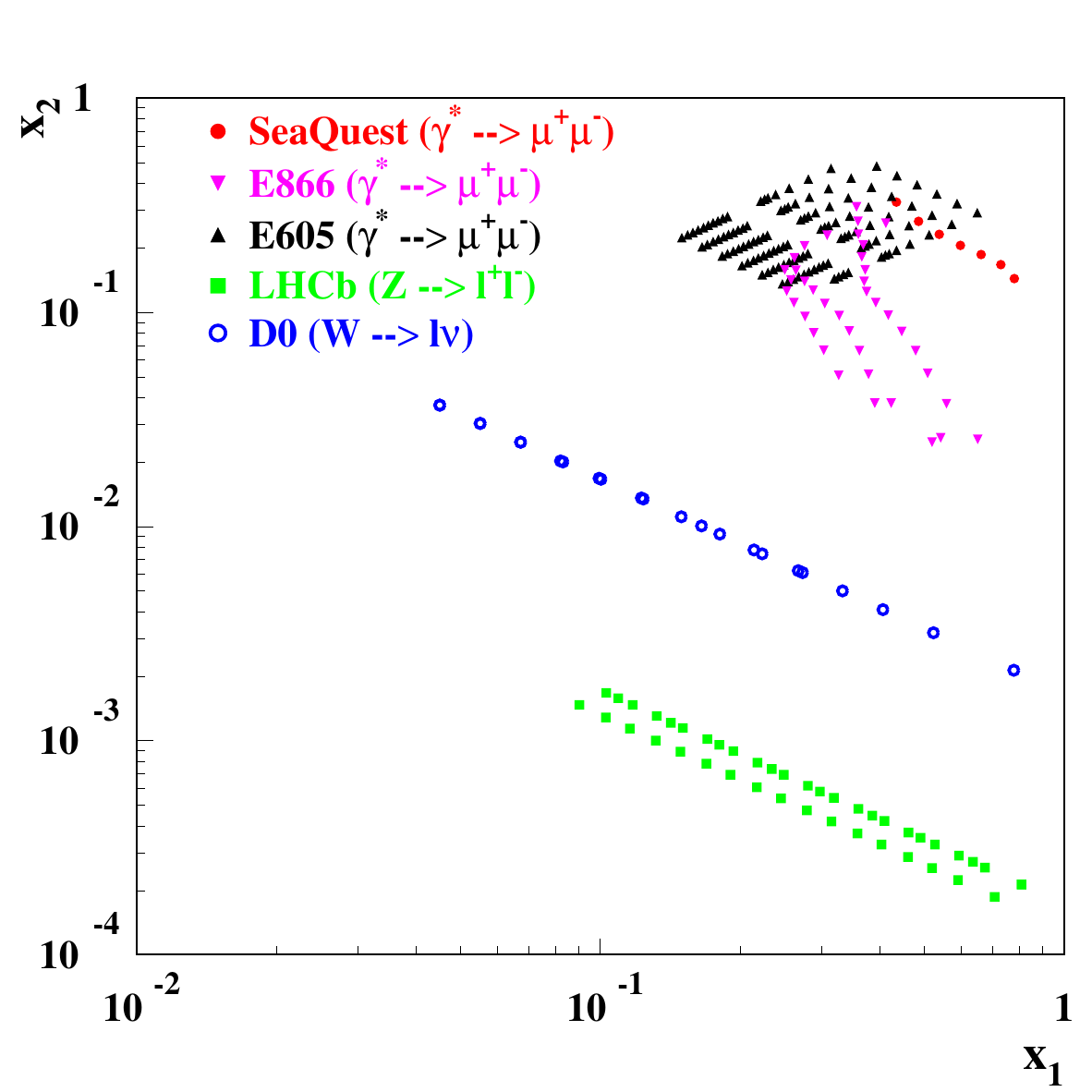}}
  \caption{\label{fig:coverage}
    The $(x_1, x_2)$ coverage of various DY datasets from collider and fixed-target experiments, used in the ABMP16~+~SeaQuest PDF fit (red full circles: SeaQuest~\cite{SeaQuest:2022vwp}, magenta down-oriented triangles: NuSea~\cite{Towell:2001nh}, black up-oriented triangles: FNAL-E605, blue open circles: Tevatron D0 $W^\pm$ production~\cite{Abazov:2013rja,D0:2014kma}, green: LHCb $Z$-production at $\sqrt{s}=7$ and 8 TeV~\cite{Aaij:2015gna,Aaij:2015vua,Aaij:2015zlq}).}   
\end{figure}

The knowledgde of parton distribution functions (PDFs) at large $x$ is one of the most urgent open issues, that, if not solved, will be capable of compromising precision physics within the Standard Model (SM) and searches for physics beyond the SM at the forthcoming high-luminosity Runs at the Large Hadron Collider (HL-LHC).

In this contribution we focus on quark distributions. Deep-inelastic-scattering (DIS) data at HERA and in fixed-target experiments allow to constrain quark distributions in a wide range of longitudinal momentum fraction $x$ values. In order to distinguish the sea and the valence quark component in case of the up and down flavour, they are supplemented with Drell-Yan (DY) data at colliders and in fixed-target experiments.

The present study aims at investigating the relevance of recently released FNAL-E906 (SeaQuest) data on the ratio of DY cross sections for proton-deuteron ($pd$) and proton-proton ($pp$) collisions $\sigma_{pd \rightarrow \gamma^* \rightarrow \mu^+\mu^-}/\sigma_{pp \rightarrow \gamma^* \rightarrow \mu^+\mu^-}$~\cite{SeaQuest:2022vwp, SeaQuest:2021zxb} as a function of Feynman-$x$ variable $x_F$ for constraining the light-flavour isospin asymmetry $(\bar{d}-\bar{u})(x)$, and the related $(\bar{d}/\bar{u})(x)$ ratio, according to an idea already pointed out by Ellis and Stirling in Ref.~\cite{Ellis:1990ti}. Neglecting nuclear effects and considering a fixed-target experiment in forward ki\-ne\-ma\-tics like SeaQuest, $\sigma_{pd \rightarrow \gamma^* \rightarrow \mu^+\mu^-}/\sigma_{pp \rightarrow \gamma^* \rightarrow \mu^+\mu^-}(x_2)$ is directly related to $(\bar{d}/\bar{u})(x_2)$, where $x_2$ is the longitudinal momentum fraction of the parton in the target nucleon.
The SeaQuest center-of-mass energy $\sqrt{s} = 15.1$ GeV is smaller than for previous experiments also sensitive to the $(\bar{d}-\bar{u})(x)$ asymmetry  and the $(\bar{d}/\bar{u})(x)$ ratio, CERN-NA51~\cite{NA51:1994xrz} and FNAL-E866 (NuSea)~\cite{Towell:2001nh}. Therefore SeaQuest can be considered as complementary to these old experiments, probing an $x$ range extending to higher values, i.e. $0.12 \lesssim x \lesssim 0.45$. On the other hand, the data from the older experiments, in particular NuSea, are essential to constrain the aforementioned isospin asymmetry at lower $x$ values, down to $x\sim 2 \cdot 10^{-2}$. Further DY data were collected by the FNAL-E605 experiment~\cite{Moreno:1990sf}, reaching an $x$ value even larger than SeaQuest, i.e. $x \sim 0.5$, however using only a Cu target. The latter are not particularly sensitive to the isospin asymmetry considering that copper is close to isoscalarity and that data with a H target were not collected by the latter experiment. 

SeaQuest released its data on DY cross-section ratio in seven bins in $x_F$. For each bin the average value of the invariant mass and transverse momentum of the $\mu^+\mu^-$ pair, $\langle M(\mu^+,\mu^-) \rangle$ and
$\langle P_T(\mu^+\mu^-) \rangle$ are also reported. With this information, one can reconstruct a value of $\langle p_L(\mu^+,\mu^-) \rangle$ and
$\langle E(\mu^+,\mu^-) \rangle$, and from there an average rapidity $\langle y(\mu^+,\mu^-) \rangle$ and, assuming leading order kinematics, an ($x_1$, $x_2$) combination of values corresponding to each data point. The latter are displayed in Fig.~\ref{fig:coverage} together with the values of ($x_1$, $x_2$) for each datapoint of other DY measurements already used in the ABMP16 PDF fit.
One can also observe, that in comparison to previous fixed-target experiments FNAL-E605 and FNAL-E866, the correlation between $x_1$ and $x_2$ values in case of SeaQuest is more evident, because the latter experiment restricted $M(\mu^+,\mu^-)$ to a small range of values $\sim 5$ GeV. A cut $M(\mu^+,\mu^-) > 4.5$ GeV was imposed to suppress the contamination from $J/\psi$ and $\psi^\prime$ production and decay. One can also observe the complementarity with collider data, where the correlation between $x_1$ and $x_2$ is more evident due to the exchange of an heavy boson ($W^\pm$ or $Z$). The latter are capable of constraining PDFs at smaller $x_2$ values at a larger scale $Q^2$.

\section{Comparison of fits incorporating or not SeaQuest data}
\label{sec:fit}

All fits discussed in this section were performed following the ABMP16
methodology discussed in Ref.~\cite{Alekhin:2017kpj}.
We considered and compare the original ABMP16 NNLO and NLO fits~\cite{Alekhin:2017kpj, Alekhin:2018pai} with
new NNLO and NLO fits including all data already in ABMP16 plus the
SeaQuest ones discussed above.
The corresponding $\chi^2$ are reported in Table~\ref{tab:datatot}.
The $\chi^2$ values per data point when including or not SeaQuest data in the fits are comparable and compatible within statistical fluctuations, meaning that adding SeaQuest data does not worsen the quality of the fit. On the other hand, the $\chi^2$ values per data point for the NNLO analyses are closer to one than those for the NLO analyses, as expected, considering that including higher-order corrections in theoretical predictions leads to a better accuracy  and a better agreement with the experimental data. 
We thus focus mainly on the NNLO analysis, and besides the two variants above, we also consider other two: one with Tevatron data excluded and the other with LHCb data excluded. The four variants correspond to the four columns of Table~\ref{tab:datahq}. For each variant, we report explicitly the $\chi^2$ for various experimental data sets. By comparing column (I) and (II), it is clear that adding the SeaQuest data does not compromise the description of other datasets, i.e. the SeaQuest data turn out to be compatible with both the NuSea and the collider ones.
On the other hand, by comparing (II) with (III) and (IV) it is evident that eliminating $W^\pm$ Tevatron data allows to improve the description of LHCb data and viceversa, i.e. a tension between the considered Tevatron and LHC data is observed. 
\begin{table}[t!]
\begin{center}                   
  \begin{tabular}{|c|c|c|c|}
\hline
fit & $\mathrm{NDP}$ & \multicolumn{2}{c|}{$\chi^2$}
\\
\cline{3-4}
 & & 
NLO & NNLO \\
\hline
ABMP16 & 2861 & 3428.9  &  3377.6
\\
\hline
ABMP16 + SeaQuest & 2868 & 3438.4  &  3384.7
\\
\hline
 \end{tabular}
\caption{\small 
\label{tab:datatot}
{The total values of $\chi^2$ obtained for the NLO and NNLO ABMP16 (+ SeaQuest) fits. See Ref.~\cite{Alekhin:2023uqx} for more detail.}}
\end{center}
\end{table}

\begin{table}[!]
\begin{center}                   
  \begin{tabular}{|l|c|c|c|c|c|c|c|c|}
    \hline
Experiment & Process & $\sqrt{s}$ (TeV) & Ref.      & $\mathrm{NDP}$& \multicolumn{4}{c|}{$\chi^2$} \\
\cline{6-9}
 &  &  &  &  & 
I & II & III & IV \\ 
\hline
SeaQuest &$pp \rightarrow \gamma^* X \rightarrow \mu ^+ \mu^- X$ &0.0151 & \cite{SeaQuest:2022vwp} & 7 & --& 7.3& 8.1 & 7.6 \\
   & $pd \rightarrow \gamma^* X \rightarrow  \mu ^+ \mu^- X$ &  &  &  &  &  & &
\\
\hline
NuSea &$pp \rightarrow \gamma^* X \rightarrow \mu ^+ \mu^- X$ &0.0388 & \cite{Towell:2001nh} & 39 & 52.8& 54.3&52.5 &53.0
\\
   &$pd \rightarrow \gamma^* X \rightarrow  \mu ^+ \mu^- X$ &  &  & & & & &
\\
\hline
D0 &$\bar{p}p  \rightarrow W^{\pm} X\rightarrow \mu ^{\pm} \overset{(-)}{\nu}  X$ &1.96& \cite{D0:2013xqc} & 10 & 17.6& 17.6&--&14.5
\\
\cline{2-9}
 &$\bar{p}p  \rightarrow W^{\pm} X\rightarrow e ^{\pm} \overset{(-)}{\nu}  X$ &1.96& \cite{D0:2014kma} & 13 & 19.0& 19.0&--&15.9
\\
\hline
LHCb &$pp  \rightarrow W^{\pm} X\rightarrow \mu ^{\pm} \overset{(-)}{\nu}  X$ &7& \cite{LHCb:2015okr} & 31 & 45.1& 43.9&35.0 &--
\\
   &$pp  \rightarrow Z X\rightarrow \mu ^{+} \mu ^{-} X$ &  &  &  &&&&
\\
\cline{2-9}
 &$pp  \rightarrow W^{\pm} X\rightarrow \mu ^{\pm} \overset{(-)}{\nu}  X$ &8& \cite{LHCb:2015mad} & 32 &40.0 &39.6 &38.2&--
\\
   &$pp  \rightarrow Z X\rightarrow \mu ^{+} \mu ^{-} X$ &  &  &  &&&&
\\
\cline{2-9}
&$pp  \rightarrow Z X\rightarrow e ^{+} e ^{-} X$ &8& \cite{LHCb:2015kwa} & 17 &21.7 &21.9 &21.9&--
\\
\hline
 \end{tabular}
\caption{\small 
\label{tab:datahq}
{The values of $\chi^2$ obtained for the data sets probing the large-$x$ light quark PDFs, included in various QCD analysis variants (column I: NNLO ABMP16 fit~\cite{Alekhin:2017kpj}, column II: NNLO ABMP16~+~SeaQuest fit, column III: a variant of II with the D0 DY data excluded, column IV: a variant of II with the LHCb DY data excluded). See Ref.~\cite{Alekhin:2023uqx} for more detail.
}}
\end{center}
\end{table}

\begin{figure}[h!]
\centerline{
    \includegraphics[width=0.45\textwidth]{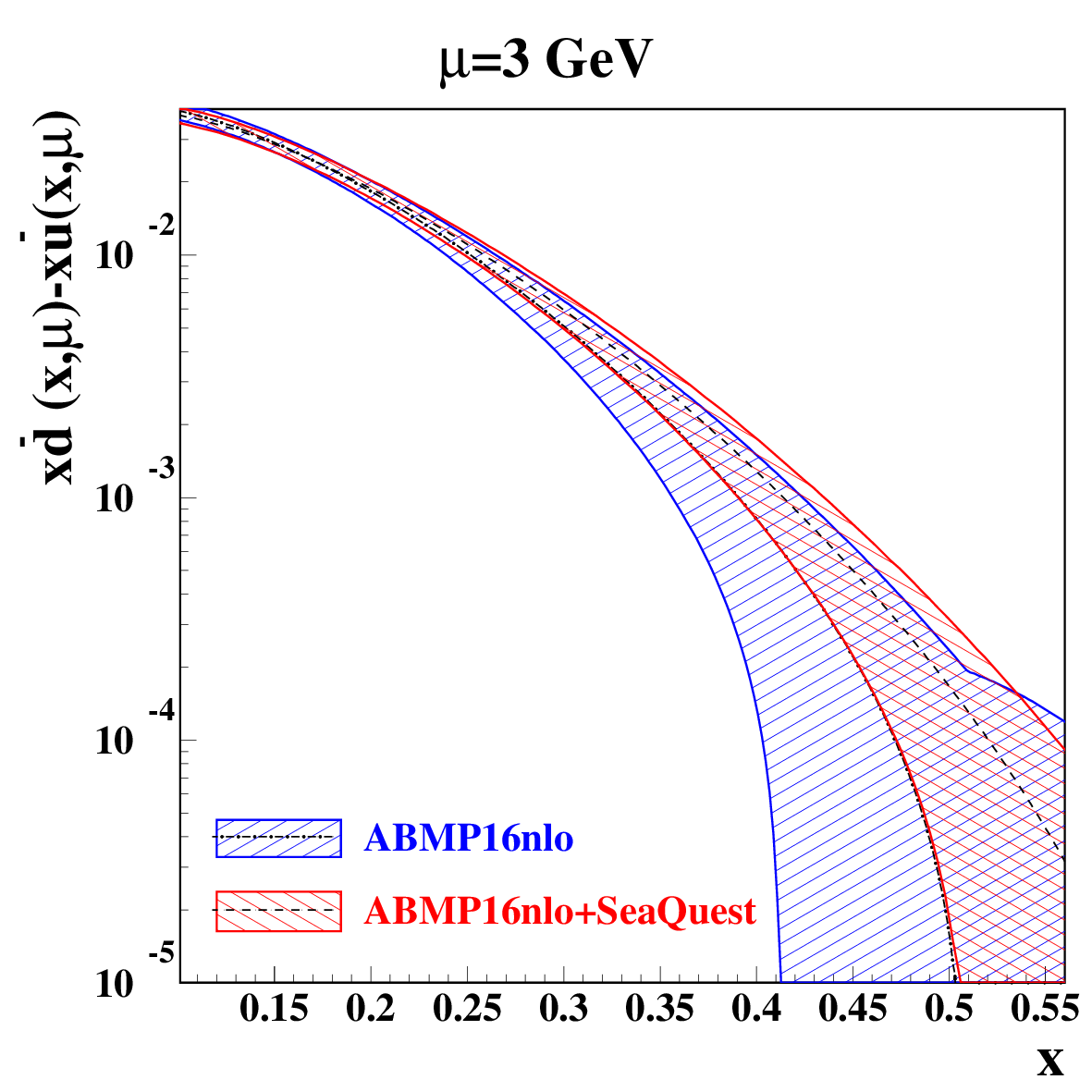}
    \includegraphics[width=0.45\textwidth]{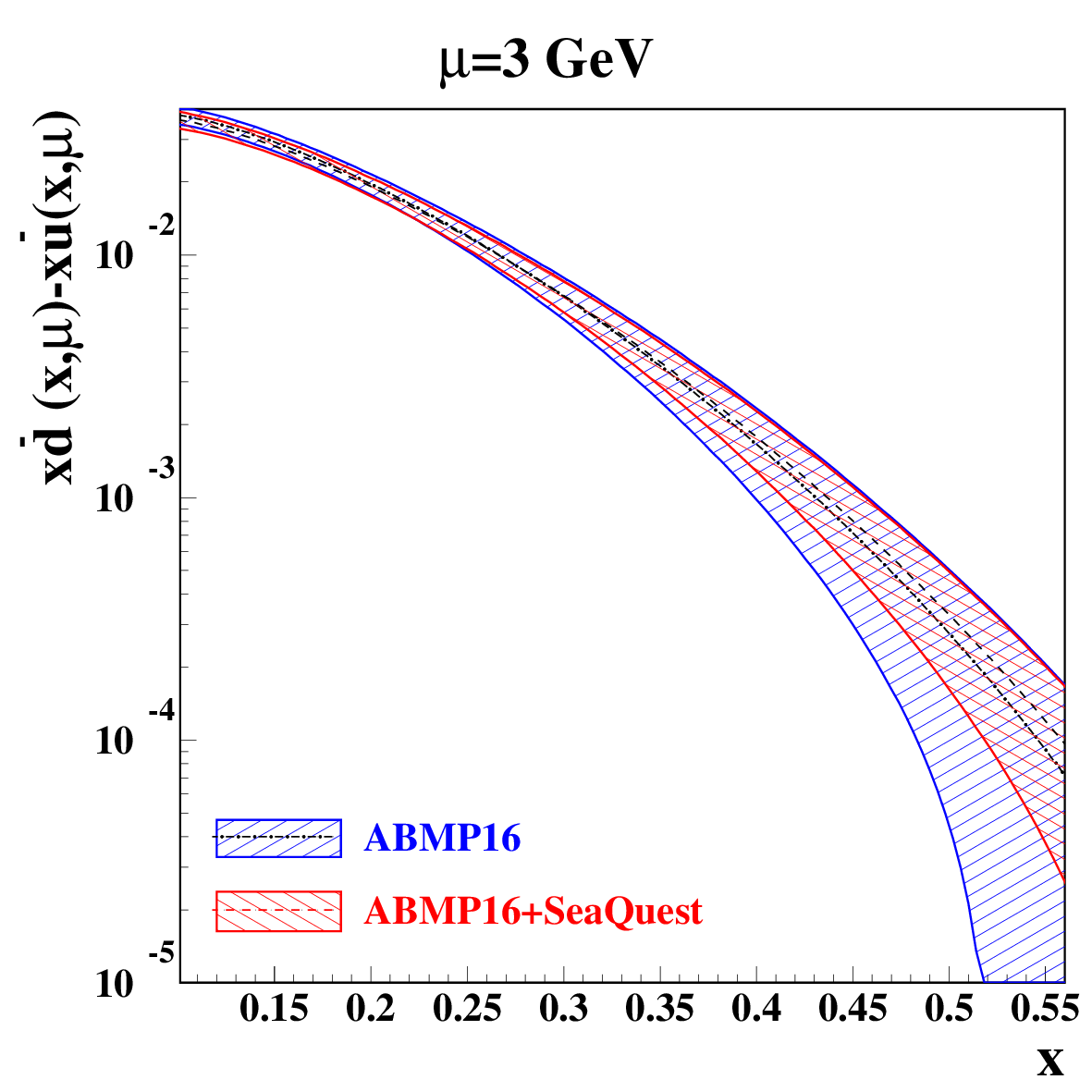}
}
\vspace*{-2mm}
  \caption{\small
  \label{fig:udm}
The $1\sigma$ band for the $n_f=3$-flavour isospin asymmetry of the sea distribution $x(\bar{d}-\bar{u}) (x)$ at the scale $\mu=$3 GeV in the ABMP16+SeaQuest analysis (central pdf: dashed line, uncertainty band: left-tilted hash) compared to the one of the ABMP16 fit (central pdf: dot-dashed line, uncertainty band: right-tilted hash). The left panel refers to the NLO analysis, whereas the right panel refers to the NNLO one.}
\end{figure}

The effect of SeaQuest data in constraining the isospin asymmetry is illustrated in Fig.~\ref{fig:udm}. It is evident that the $1\sigma$ uncertainty band on the latter shrinks for $x > 0.3$ in the ABMP16~+~SeaQuest analysis in comparison to the ABMP16 one. However, one should also note that at large $x$ the isospin asymmetry become increasingly small. These observations apply to both the NLO and the NNLO fit.

\section{Nuclear effects}
Considering that SeaQuest data involve the use of a deuteron target, besides a proton one, we investigated the role of nuclear effects~\cite{Alekhin:2023uqx}. We found them small, as evident when comparing the ratio of $\sigma_{pd \rightarrow \gamma^* \rightarrow \mu^+\mu^-}$ computed accounting for the fact that partons from the target are bound in a nucleus, to the sum of $\sigma_{pp \rightarrow \gamma^* \rightarrow \mu^+\mu^-}$ and $\sigma_{pn \rightarrow \gamma^* \rightarrow \mu^+\mu^-}$, which neglects nuclear effects. As shown in Ref.~\cite{Alekhin:2023uqx}, the ratio vary between 1.005 and 1.01 in the SeaQuest bins $x_F > 0.2$, while it increases when considering smaller $x_F$ (corresponding to larger $x_2$) values outside of the SeaQuest present coverage. We concluded that nuclear effects due to deuteron can be neglected when using SeaQuest data in proton PDF fits at the current level of data uncertainties.

\section{Conclusions}
\label{sec:conclu}

In summary, we found that SeaQuest data are capable of reducing the uncertainties on the $(\bar{d}-\bar{u})(x)$ asymmetry at large $x$, in particular $0.3 < x < 0.45$ and their role is complementary to those of the NuSea data at smaller $x$. The constraints from SeaQuest and NuSea data are compatible with those from collider data. The results point towards an asymmetric sea, with a predominance of $\bar{d}(x)$ quarks vs. $\bar{u}(x)$ at $x \sim 0.3$. 

So far the SeaQuest collaboration has analyzed only a part of the data they already collected. We expect that completing the analysis will be fundamental in order to decrease the present data uncertainties $\sim 5\%$ and to increase their constraining power. 

The ABMP16 + SeaQuest NNLO PDF fit will soon be released in the LHAPDF library~\cite{Buckley:2014ana}.

\section*{Acknowledgements}
The work of S.A. and S.-O.M. has been supported in part by  the Deutsche Forschungsgemeinschaft through the Research Unit FOR 2926, {\it Next Generation pQCD for Hadron Structure: Preparing for the EIC}, project number 40824754. The work of M.V.G. and S.-O.M. has been supported in part by the Bundesministerium f\"ur Bildung und Forschung under contract 05H21GUCCA.

\bibliographystyle{JHEP}
\bibliography{seaquestproc}

\end{document}